\documentclass[]{spie}  

\usepackage{amsmath,amsfonts,amssymb}
\usepackage{graphicx}
\usepackage{comment}
\usepackage[colorlinks=true, allcolors=blue]{hyperref}

\title{A Comparative Study of 2D Image Segmentation Algorithms for Traumatic Brain Lesions Using CT Data from the ProTECTIII Multicenter Clinical Trial}

\author[*a]{Shruti Jadon M.S.}
\author[a,b] {Owen P. Leary Sc.B.}
\author[a]{Ian Pan M.S.}
\author[a]{Tyler J. Harder M.S.}
\author[d]{David W. Wright MD}
\author[a,b,c,e]{Lisa H. Merck M.D. M.P.H.}
\author[a,e,f]{Derek L. Merck Ph.D.}
\affil[a]{Warren Alpert Medical School of Brown University, Diagnostic Imaging,Providence RI 02903}
\affil[b]{Warren Alpert Medical School of Brown University, Neurosurgery,Providence RI 02903}
\affil[c]{Warren Alpert Medical School of Brown University, Emergency Medicine, Providence RI 02903}
\affil[d]{Emory University School of Medicine, Department of Emergency Medicine, Atlanta GA 30322.}
\affil[e]{University of Florida School of Medicine, Department of Emergency Medicine, Gainesville FL 32611.}
\affil[f]{Brown University Department of Engineering, Providence RI 02906.}

\authorinfo{* visiting researcher at Rhode Island Hospital, Department of Diagnostic Imaging.
\\ Code Available at Github: \url{https://github.com/shruti-jadon/Brain-Lesions-Segmentation}}

\begin{document} 
\maketitle

\begin{abstract}
Automated segmentation of medical imaging is of broad interest to clinicians and machine learning researchers alike. The goal of segmentation is to increase efficiency and simplicity of visualization and quantification of regions of interest within a medical image. Image segmentation is a difficult task because of multiparametric heterogeneity within the images, an obstacle that has proven especially challenging in efforts to automate the segmentation of brain lesions from non-contrast head computed tomography (CT). In this research, we have experimented with multiple available deep learning architectures to segment different phenotypes of hemorrhagic lesions found after moderate to severe traumatic brain injury (TBI). These include: intraparenchymal hemorrhage (IPH), subdural hematoma (SDH), epidural hematoma (EDH), and traumatic contusions. We were able to achieve an optimal Dice Coefficient\cite{article1} score of 0.94 using UNet++ 2D Architecture with Focal Tversky Loss Function, an increase from 0.85 using UNet 2D with Binary Cross-Entropy Loss Function in intraparenchymal hemorrhage (IPH) cases. Furthermore, using the same setting, we were able to achieve the Dice Coefficient score of 0.90 and 0.86 in cases of Extra-Axial bleeds and Traumatic contusions, respectively.    
\end{abstract}


\section{Background}
\label{sec:intro}  

Traumatic brain injury (TBI)\cite{article1} \cite{article2} is a leading cause of morbidity and mortality \cite{article} in children and young adults in the United States \cite{article7}. TBI is also a major concern for elderly individuals, with high rates of mortality and hospitalization due to falls among people aged 75 and older. Depending on the severity of injury\cite{Chong2015PredictiveMI}, TBI can have a lasting impact on quality of life for survivors of all ages. Long-term deficits may include impairments of cognition, decision making, concentration, memory, movement, or sensation (vision or hearing), as well as emotional problems (personality changes, impulsivity, anxiety, and depression)\cite{Tepas2009} and epilepsy \cite{article7}.Annually, TBI injuries cost an estimated \$76 billion in direct and indirect medical expenses in the US\cite{article7}. The long-term effects of TBI vary depending on patient demographics and availability of prompt medical care at a trauma center, among other factors.  \\
While rapid diagnosis and treatment of TBI is directly related to patient outcomes, attempts to automate identification\cite{vedaldi2014matconvnet}, classification, and segmentation\cite{10.3389/fneur.2019.00541} \cite{article4} of traumatic brain lesions on non-contrast head computed tomography (CT), the gold standard for emergent imaging diagnosis of TBI, have been met with limited success. Open research on machine learning resources for TBI has been constrained by few large public patient imaging datasets, as well as the emphasis of corporate developers on proprietary algorithms to be incorporated into for-profit software packages. Despite such limitations, a few studies have attempted classification and/or detection of TBI on CT and magnetic resonance (MR) imaging, including concussion classification\cite{Mitra2016StatisticalML} using implicit features of the entire voxel-wise white matter fiber strains and detection of brain lesions\cite{Mitra2016StatisticalML}.One group using Convolutional Neural Networks (CNNs) and Scale-Invariant Feature Transform (SIFT , edge detection algorithm)\cite{2017SPIE10134E..2GK} on 409 CT scans was able to achieve 92.55\% detection accuracy [8]. Other groups have attempted prediction of mortality rate after traumatic brain injury using logistic regression-based models\cite{2017SPIE10134E..2GK}, probability-based models (Naïve Bayes)\cite{2018PLoSO..1307192R}, Support Vector Machine, and Artificial Neural Network (ANN). Maximum accuracies achieved included 81-84\% accuracy using regression, and ~84\% using ANN\cite{2018PLoSO..1307192R}. So far, less work has been done in automated volumetric segmentation\cite{Juang2010MRIBL}, mainly due to lack of large annotated datasets. A group\cite{2018arXiv180710839R} using inception-based convolutional neural network on dataset of 18 patients and were able to achieve dice score of 0.75, and others working on MR FLAIR\cite{10.3389/fneur.2019.00541},\cite{article4} stroke lesion segmentation using a customized MRF (markov random fields) method and U-Net model respectively, and were able to achieve median Dice Coefficient of ~0.79.   \\
Application and comparison of available algorithms for rapid segmentation of traumatic lesions is important for translating more accessible, fully automated\cite{Phan2018AutomaticDA} diagnostic tools\cite{Roy2015AUTOMATEDBH}, \cite{Fedorov20123DSA},\cite{Gillebert2014} to the clinic. Automated segmentation (AS) can be used to increase efficiency and precision of pathology visualization and quantification on non-contrast head CT. However, intra-image and intra-lesion heterogeneity across multiple feature parameters, such as variable Hounsfield unit ranges and irregular lesion boundaries, can pose technical challenges to the automatic segmentation of traumatic brain lesions\cite{Mitra2016StatisticalML}\cite{Juang2010MRIBL}. In this work, we applied and compared previously published deep learning architectures (e.g. SegNet, DenseNet, U-Net, etc) for semantic segmentation of different classes of traumatic brain lesions, including solitary intraparenchymal hemorrhage (IPH), subdural hematoma (SDH), epidural hematoma (EDH), and multifocal hemorrhagic contusions. Semantic segmentation is a method for assigning a class label to each pixel in a two-dimensional (2D) image, and thereby producing 2D lesion segmentations on each slice positive for pathology of interest. Models were tested using a large CT imaging dataset collected during the ProTECTIII multicenter clinical trial\cite{doi:10.1080/10903127.2017.1315201} of intravenous progesterone vs. placebo for moderate to severe TBI, as well as associated ground truth slice-by-slice volumetric annotations of lesions of interest. Models were compared in terms of Dice Coefficient\cite{article1} (DC), Sensitivity\cite{Parikh2008}, and specificity\cite{Parikh2008}.

\section{Method}
In this work, we selected open source algorithms previously applied for three-dimensional (3D)\cite{Kamnitsas2017EfficientM3} and 2D segmentation of traumatic brain lesions\cite{Mitra2016StatisticalML} on NCCT on the basis of their documented performance on existing databases (such as the COCO dataset\cite{Lin2014MicrosoftCC}) for semantic segmentation. Our dataset (refer to section 2.1) was grouped according to index lesion phenotype, including solitary intraparenchymal hemorrhage (IPH), multifocal hemorrhagic contusion, and subdural / epidural hematoma (SDH \& EDH, grouped together as “extra-axial bleeds”). Algorithms were applied separately to data in each lesion phenotype category. Algorithms were compared across bleed phenotypes in terms of dice coefficient\cite{article1}, sensitivity\cite{Parikh2008}, and specificity\cite{Parikh2008} in two phases. First, 2D and 3D segmentation algorithms were compared in terms of accuracy, with 2D algorithms tested across single slice data points and 3D algorithms tested across single case data points. Based on preliminary data from both groups of algorithms suggesting poorer performance of 3D methods, only 2D methods were then compared against each other in terms of the same parameters. Apart from segmentation modeling variation, we also experimented with various objective functions for sparse segmentation. The project code has been made available for validation and replication of these experiments and can be found at \url{https://github.com/shruti-jadon/Brain-Lesions-Segmentation}.

\subsection{Imaging Dataset \& Pre-Processing}
\label{sec:title}
881 patients with clinical impression of moderate-severe TBI were enrolled in the ProTECTIII clinical trial (NCT00822900) between 2010 and 2014 and underwent baseline head CT imaging to identify intracranial\cite{Chen2010IntracranialPL} pathology. Imaging was carried out across 22 trauma centers in the United States according to site-specific CT imaging protocols \cite{Wright2014VeryEA}. After collection of all imaging, a central neuroradiologist screened and identified all baseline CTs with the largest lesion classified as one of four pre-specified bleed phenotypes of interest (SDH, EDH, multifocal hemorrhagic contusion, or solitary IPH). Overall, 516 baseline CT scans of unique study participants, each with approximately 100 slices, were selected for analysis. All imaging data were anonymized and stored in a central imaging repository. Local IRB approval was obtained for the project at the institution were imaging analyses were performed. \\
Trained research associates annotated all 516 head CTs in standardized fashion using 3D Slicer (Version 4, Brigham \& Women’s Hospital, 2012) to trace the boundary of the largest identified qualifying lesion (i.e. one of the pre-specified bleed phenotypes) to produce one volumetric annotation per case. In 17 of the 516 cases, two lesions were segmented, yielding a total of 533 unique lesions. The resultant ground-truth dataset included 43 IPH annotations, 48 EDH annotations, 212 SDH annotations, and 230 hemorrhagic contusion annotations.  From these data, individual 2D slices of each patient’s baseline non-contrast head CT scan were considered as individual instances for algorithmic experiments. After data segregation, we also applied image preprocessing and normalization technique using nilearn library\cite{DBLP:journals/corr/AbrahamPEGMKGTV14}.


   \begin{figure} [!ht]
   \begin{center}
   \begin{tabular}{c} 
   \includegraphics[height=5cm]{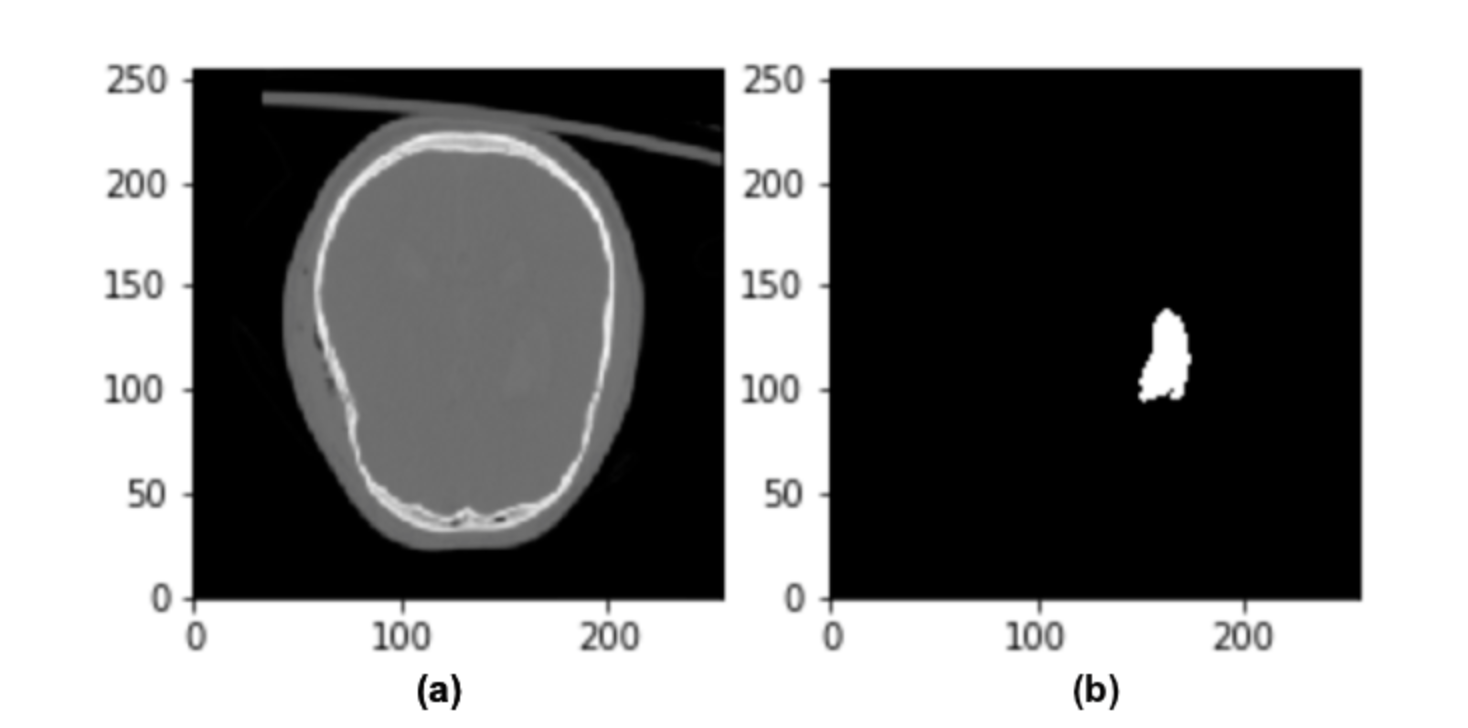}
   \end{tabular}
   \end{center}
   \caption[example] 
   { \label{fig:example} 
Sample TBI case from the ProTECTIII dataset\cite{doi:10.1080/10903127.2017.1315201} demonstrating NCCT with bone windowing and level (a) and outline of ground truth temporal lobe intraparenchymal hemorrhage annotation (b).}
   \end{figure}

\subsection{Segmentation Algorithms}
Medical image segmentation is one of the most challenging tasks in computer vision. For an approach to be successful, it should employ methods capable of covering all the outlying cases (e.g. those with sparse segmentation). Traumatic brain lesion segmentation is particularly challenging due to the small number of image features which are ubiquitously present across individual cases and diverse bleed phenotypes. Unsurprisingly, experiments with existing UNet-3D\cite{iek20163DUL} models met limited success in our preliminary analysis, likely due to the relatively small data set (i.e. 533 data points) for discerning over 1.3 million parameters amidst significant inter-case feature heterogeneity.
Higher capacity models capable of segregating skull from brain, and data sets including more case examples of each lesion phenotype across various brain locations, are probably necessary to improve 3D segmentation performance\cite{Jadon2019ImprovingSN}. On the other hand, the 2D segmentation model leverages a greater number of data points from the same dataset\cite{jadon2019hands}; we have approximately 100 slices per CT scan, comprising a total of approximately 58000 input data points. Based on preliminary results, our analytic comparison focused on 2D segmentation models. In particular, we explored multiple algorithms across two categories of machine learning 2D modeling techniques: 
\begin{itemize}
    \item Image Segmentation models (U-Net 2d; Inception based UNet++ 2D model)
    \item Objective Functions (Binary Cross-Entropy\cite{Zhang2018GeneralizedCE}; Dice Loss\cite{Sudre2017GeneralisedDO}; Focal Loss; Focal Tversky Loss).  
\end{itemize} 

  \begin{figure} [!ht]
   \begin{center}
   \begin{tabular}{c}
   \includegraphics[height=6cm]{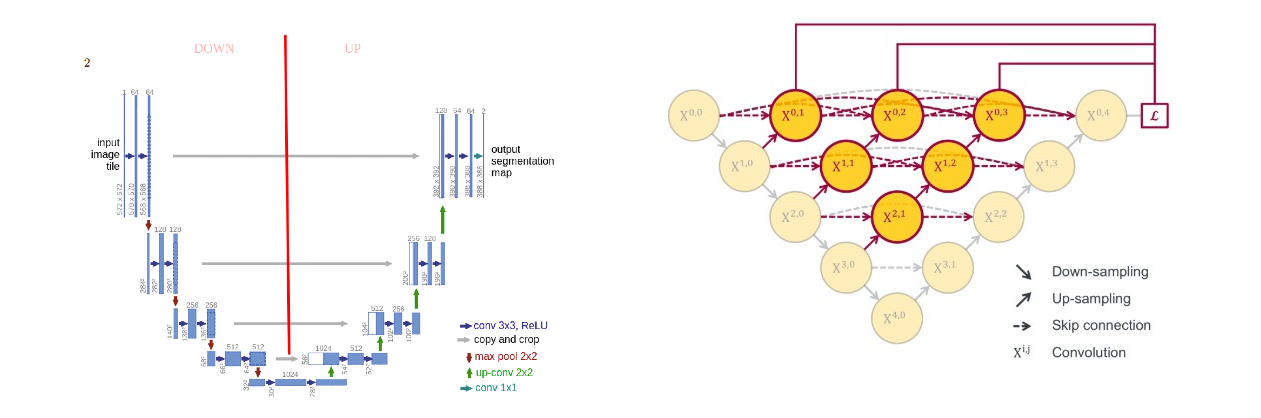}
   \end{tabular}
   \end{center}
   \caption[example] 
   { \label{fig:example} 
Sample UNet2D\cite{Ronneberger2015UNetCN} and  UNet++ 2D \cite{Zhou2018UNetAN} layer architecture}
   \end{figure} 
   
UNet\cite{Ronneberger2015UNetCN} is a convolutional neural network model designed for biomedical image segmentation. The architecture contains two parts: an encoder (compression)\cite{jadon2018introduction} consisting of a stack of convolutional and max pooling layers used to capture the context of the image, and a decoder (expansion) which enables precise localization using transposed convolutions. UNet is therefore is a fully convolutional network (FCN), i.e. it only contains convolutional layers. UNet++ 2D model differs from standard UNet2D\cite{Ronneberger2015UNetCN} in that it’s type of layers’ and architecture. UNet++ 2D combines initial layer features with all preceding layers, making it inception based across whole network, this leads to a densely connected encoder-decoder architecture. This helps the model to keep the context of input data across all layers, and thereby helps converging model faster.

\subsection{Objective Functions}
Objective functions are also known as a “loss functions,” and also play a crucial role in determining the success of a model. Objective functions map required information from data into space where we can measure cost and attempt to reduce it using some optimization algorithm such as stochastic gradient descent\cite{Ruder2016AnOO}. There are various forms of objective functions that can be used for semantic segmentation. The four used in this paper are summarized subsequently.

\subsubsection{Binary Cross-Entropy (BCE)}
Cross-Entropy is a method to map information of data into categorical distribution. As name suggests, Binary Cross-Entropy\cite{Zhang2018GeneralizedCE} means 2 class categorical distribution.
\begin{align*}
  CrossEntropy(p,\hat{p}) &= -(p*log(\hat{p})+(1-p)log(1-\hat{p}))
\end{align*}
\subsubsection{Dice Loss}
The Dice Coefficient\cite{article1} is similar to the Jaccard Index (Intersection over Union, IoU), where TP are the true positives, FP false positives and FN false negatives. The Dice Coefficient\cite{article1} can also be defined as a loss function: 

\begin{align*}
  Dice Loss(p,\hat{p}) &= 1-\frac{2p \hat{p}}{p+\hat{p}+1}
\end{align*}

\subsubsection{Focal Loss}
Focal Loss\cite{lin2017focal} (FL) attempts to down-weight the contribution of easy examples so that the CNN focuses more on hard examples. Focal Loss\cite{lin2017focal} can be defined as follows: 

$FocalLoss(p,\hat{p})=-(\alpha(1-\hat{p})^\gamma p log(\hat{p})+ (1-\alpha)\hat{p}^\gamma (1-p)log(1-\hat{p}))$

\subsubsection{Focal Tversky Loss} Tversky index (TI) is a generalization of Dice’s coefficient. TI adds a weight to FP (false positives) and FN (false negatives), which is just the regular Dice Coefficient\cite{article1}. Similarly, to Dice Loss\cite{Sudre2017GeneralisedDO}, the Tversky Loss function can be defined as follows: 

$ TI(p,\hat{p})=\frac{p\hat{p}}{p\hat{p}+(1-\beta) p(1-\hat{p})+\beta \hat{p}(1-p)} $
and Focal tversky 
$FTL=\sum(1 - TI_c)^\frac{1}{\gamma}$
where $\gamma$ varies in the range [1, 3]

\subsection{Evaluation Metrics}
One of the most challenging aspects of machine learning for healthcare applications is measurement of accuracy, especially when it comes to clinical acceptability of false positive and negative rates. There are various metrics to measure the performance of a model ranging from accuracy to Jaccard distance. For this work, we have chosen the following metrics as measure to compare segmentation\cite{Havaei2017BrainTS} approaches: 

\subsubsection{Dice Coefficient}
The Dice Coefficient\cite{article1} also called the overlap index, is the most used metric in validating medical volume segmentations. In addition to the direct comparison between automatic and ground truth segmentations, it is common to use the DICE to measure reproducibility (repeatability). [2] used the DICE as a measure of the reproducibility as a statistical validation of manual annotation where segmenters repeatedly annotated the same MRI image, then the pairwise overlap of the repeated segmentations is calculated using the DICE, which is defined by 
\begin{align*}
  Dice Coefficient(DC) &= \frac{2TP}{2TP+FP+FN}= \frac{2|X,Y|}{|X|+|Y|}
\end{align*}
\subsubsection{Sensitivity} 
True Positive Rate (TPR), also called Sensitivity\cite{Parikh2008} and Recall, measures the portion of positive voxels in the ground truth that are also identified as positive by the segmentation being evaluated. 

\begin{align*}
  Sensitivity(TPR) &= \frac{TP}{TP+FN}
\end{align*}
\subsubsection{Specificity}
True Negative Rate (TNR), also called specificity\cite{Parikh2008}, measures the portion of negative voxels (background) in the ground truth segmentation that are also identified as negative by the segmentation being evaluated.

\begin{align*}
  Specificity(TNR) &= \frac{TN}{TN+FP}
\end{align*}

  \begin{figure} [!ht]
  \begin{center}
  \begin{tabular}{c} 
  \includegraphics[height=12cm]{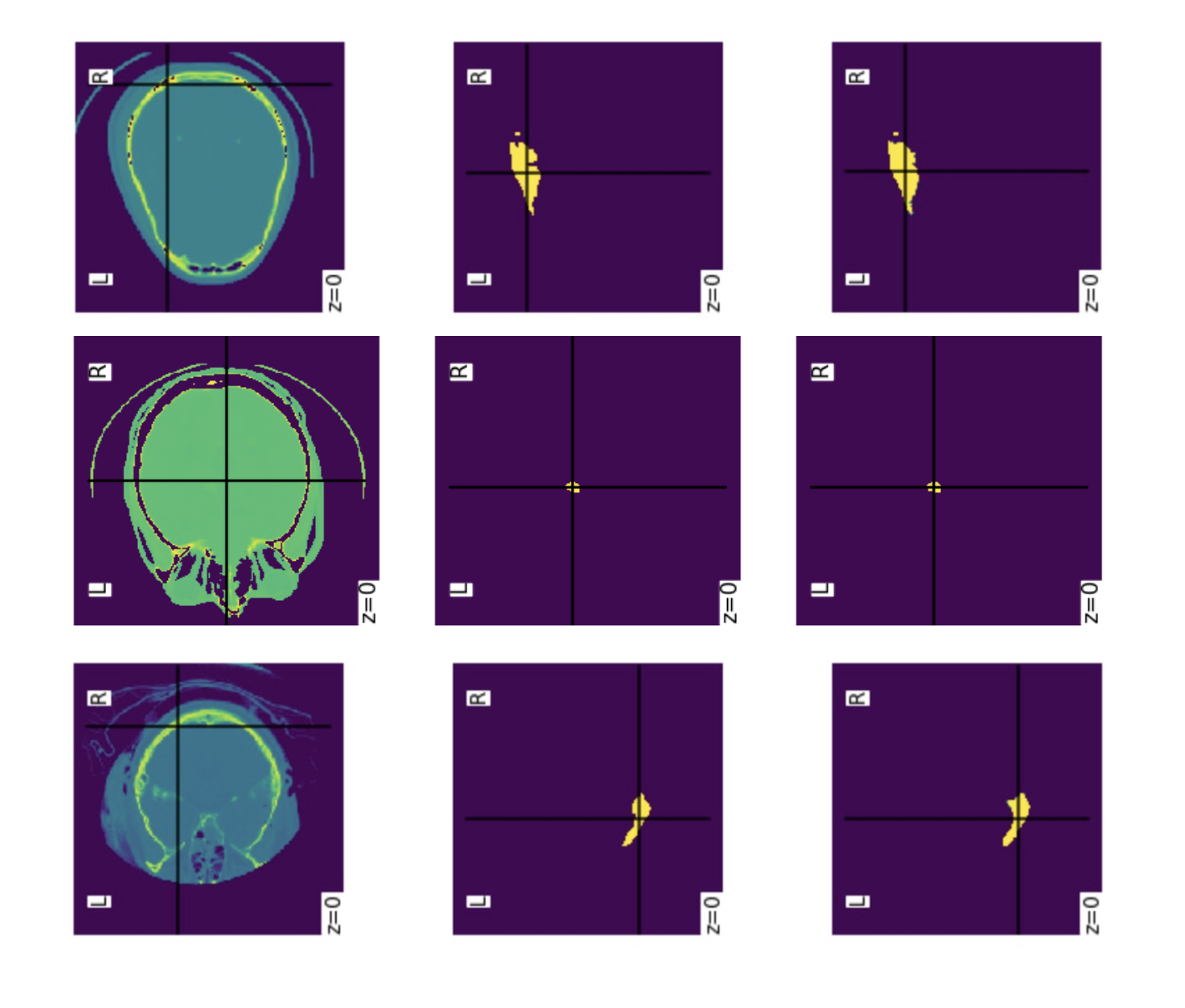}
	\end{tabular}
	\end{center}
  \caption[example] 
{ \label{fig:example} 
    Selected slices from a ProtectIII MRI dataset with lesions (Left) and corresponding mask (Center) and lesion segmentation (Right) using UNet++ 2D segmentation model with Focal Tversky Loss objective function.) 
}
  \end{figure} 

\section{Results}
3D segmentation algorithms selected for testing included only 3D-Unet. 2D segmentation algorithms tested included two image segmentation models (U-Net 2d; UNet++ 2D model) and three objective functions (Binary Cross-Entropy; Dice Loss; Focal Loss; Focal Tversky Loss). Preliminary data comparing 3D-Unet to 2D-UNet found that due to lack of data and need to train high capacity model of ~1.3 million parameters, 3D models were underfitting. Attempts to increase the number of iterations led to overfitting, with an average validation Dice Coefficient of ~0.37. These preliminary results suggested inferior performance of 3D relative to 2D algorithms on this dataset. 
2D segmentation algorithms were then compared in detail. In IPH  cases, using a total sample of 2700 CT scan 2D slices over 200 epochs, we were able to achieve an optimized Dice Coefficient score of \textbf{0.94}(refer table 1) using UNet++ 2D Architecture with Focal Tversky Loss Function, an increase from 0.85 using UNet2D with Binary Cross-Entropy Loss Function.  Similarly, we have observed an increase in Dice Coefficient score in hemorrhagic contusion from 0.79 to \textbf{0.90}(refer table 2)  by using ~13000 2D images, and in case of extra-axial bleeds (SDH \& EDH) from 0.71 to \textbf{0.85}(refer table 3)  using ~15000 images. \break



  \begin{table}[!ht]
\caption{The Dice scores, Sensitivity and Specificity for the 2 lesion segmentation methods (UNet 2D, UNet++ 2D) using 4 objective functions (Binary Cross Entropy, Dice Loss, Focal Loss, Focal Tversky Loss) on Intraparenchymal Hemorrhage (IPH) Data} 
\label{tab:Multimedia-Specifications1}
\begin{center}       
\begin{tabular}{|l|l|l|l|l|}
\hline
\rule[-1ex]{0pt}{3.0ex}  Architecture & Objective Function & Dice Coefficient & Sensitivity & Specificity  \\
\hline
\rule[-1ex]{0pt}{3.5ex} UNet 2D & Binary Cross Entropy & 0.8485 & 0.8315 & 0.9996
    \\
\hline
\rule[-1ex]{0pt}{3.5ex}  UNet 2D & Focal Loss & 0.8022 & 0.8385 & 0.9996
\\
\hline
\rule[-1ex]{0pt}{3.5ex}  UNet 2D & Dice Loss & 0.7945 & 0.7240 &	0.9994 \\
\hline
\rule[-1ex]{0pt}{3.5ex}  UNet 2D & Focal Tversky Loss & 0.8667 & 0.8661 & 0.9994
  \\
\hline
\rule[-1ex]{0pt}{3.5ex} UNet++ 2D & Binary Cross Entropy & 0.8784 & 0.8571 & 0.9996

    \\
\hline
\rule[-1ex]{0pt}{3.5ex}  UNet++ 2D & Focal Loss & 0.8776 & 0.9257 & 0.9999
\\
\hline
\rule[-1ex]{0pt}{3.5ex}  UNet++ 2D & Dice Loss & 0.8377 & 0.8051 & 0.9995
\\
\hline
\rule[-1ex]{0pt}{3.5ex}  UNet++ 2D & Focal Tversky Loss & \textbf{0.9424} & \textbf{0.9449} & \textbf{0.9998}
  \\
\hline 
\end{tabular}
\end{center}
\end{table}

  \begin{table}[!ht]
\caption{The Dice scores, Sensitivity and Specificity for the 2 lesion segmentation methods (UNet 2D, UNet++ 2D) using 4 objective functions (Binary Cross Entropy, Dice Loss, Focal Loss, Focal Tversky Loss) on Hemorrhagic Contusions Data} 
\label{tab:Multimedia-Specifications2}
\begin{center}       
\begin{tabular}{|l|l|l|l|l|}
\hline
\rule[-1ex]{0pt}{3.0ex}  Architecture & Objective Function & Dice Coefficient & Sensitivity & Specificity  \\
\hline
\rule[-1ex]{0pt}{3.5ex} UNet 2D & Binary Cross Entropy & 0.7894 &	0.7860 & 0.9996
    \\
\hline
\rule[-1ex]{0pt}{3.5ex}  UNet 2D & Focal Loss & 0.5833 & 0.6458 &	\textbf{0.9998}
\\
\hline
\rule[-1ex]{0pt}{3.5ex}  UNet 2D & Dice Loss & 0.8480 &	0.8486 &	0.9996\\
\hline
\rule[-1ex]{0pt}{3.5ex}  UNet 2D & Focal Tversky Loss & 0.8413 &	0.8880 &	0.9994
  \\
\hline
\rule[-1ex]{0pt}{3.5ex} UNet++ 2D & Binary Cross Entropy & 0.7979 & 0.8109 & 0.9996
    \\
\hline
\rule[-1ex]{0pt}{3.5ex}  UNet++ 2D & Focal Loss & 0.8458 & 0.8438 & 0.9996
\\
\hline
\rule[-1ex]{0pt}{3.5ex}  UNet++ 2D & Dice Loss &0.8402
&	0.8331 &
	0.9996

\\
\hline
\rule[-1ex]{0pt}{3.5ex}  UNet++ 2D & Focal Tversky Loss  & \textbf{0.9015} & \textbf{0.9430} & 0.9996
  \\
\hline 
\end{tabular}
\end{center}
\end{table}

  \begin{table}[!ht]
\caption{The Dice scores, Sensitivity and Specificity for the 2 lesion segmentation methods (UNet 2D, UNet++ 2D) using 4 objective functions (Binary Cross Entropy, Dice Loss, Focal Loss, Focal Tversky Loss) on Extra-Axial bleeds (SDH and EDH) Data} 
\label{tab:Multimedia-Specifications3}
\begin{center}       
\begin{tabular}{|l|l|l|l|l|}
\hline
\rule[-1ex]{0pt}{3.0ex}  Architecture & Objective Function & Dice Coefficient & Sensitivity & Specificity  \\
\hline
\rule[-1ex]{0pt}{3.5ex} UNet 2D & Binary Cross Entropy & 0.7128
	 &0.7351&
	0.9994

    \\
\hline
\rule[-1ex]{0pt}{3.5ex}  UNet 2D & Focal Loss & 0.4818
	&0.6847&
	\textbf{0.9998}

\\
\hline
\rule[-1ex]{0pt}{3.5ex}  UNet 2D & Dice Loss & 0.6879
	&0.6553&
	0.9991

 \\
\hline
\rule[-1ex]{0pt}{3.5ex}  UNet 2D & Focal Tversky Loss & 0.7438	&0.7983&
	0.9986

  \\
\hline
\rule[-1ex]{0pt}{3.5ex} UNet++ 2D & Binary Cross Entropy & 0.7276	&0.7125&	0.9989

    \\
\hline
\rule[-1ex]{0pt}{3.5ex}  UNet++ 2D & Focal Loss & 0.6966	&0.7736&	0.9998
\\
\hline
\rule[-1ex]{0pt}{3.5ex}  UNet++ 2D & Dice Loss & 0.7252

	&0.7269&

	0.9989

\\
\hline
\rule[-1ex]{0pt}{3.5ex}  UNet++ 2D & Focal Tversky Loss & \textbf{0.8578} & \textbf{0.8524} & 0.9998
  \\
\hline 
\end{tabular}
\end{center}
\end{table}
\subsection{Implications, Limitations, and Future Directions}
Automated traumatic brain injury detection and segmentation has the potential to assist physicians in faster and more accurate decision-making. As demonstrated in this work, with only 516 CT scans, we were able to extract ~40,000 2D slices and train a 2D segmentation model with very high dice coefficient relative to those previously reported. 2D segmentation accuracy could likely be further improved with even larger datasets that similarly leverage multi-institutional patient cohorts and higher capacity models. In this work, we trained our models with ~800,000 parameters, as our dataset consisted of ~40,000 examples. In the past, the deep learning community has met success in developing segmentation models with the help of large amounts of data (for example, the Coco Dataset consists of ~330,000 data points). Establishing similarly large datasets in medical image analysis will be necessary to continue to drive success in automated CT segmentation. Multi-model approaches to training a single model capable of detecting and segmenting all forms of brain lesions should also be assessed using large, heterogeneous datasets.

\section{Conclusion and Future Directions}
In this paper, we have explored well known 2D segmentation architectures of UNet 2D and UNet++ 2D, along with experiments with different objective functions. Our results suggest that UNet++ 2D works better than normal UNet 2D, suggesting the importance of combination of initial layer’s extracted features with final layer’s features. Among tested loss functions, Focal Tversky loss\cite{Abraham2018ANF} worked best for single-lesion segmentation, likely reflecting the relatively small proportion of each slice by area occupied by lesion annotation. Since annotation masks are relatively small, the algorithms which assign more weight to the annotations relative to other features perform better by ~0.10 dice coefficient. Compared to previous reports which have applied algorithmic approaches to automated 2D segmentation of traumatic brain lesion, we report an improvement in Dice Coefficient\cite{article1} by approximately 0.08 for IPH and extra-axial hematomas.

\nocite{*}
\bibliography{main} 
\bibliographystyle{spiebib} 

\end{document}